\documentclass[amsmath,amssymb,aps,twocolumn]{revtex4}

\newcommand{\lp}{\left}
\newcommand{\rp}{\right}
\newcommand{\EE}{{\cal E}}
\newcommand{\OO}{{\cal O}}
\newcommand{\WT}{\widetilde}

\begin{document}


\title{Cascading DGP}

\author{Claudia~de~Rham$^{a,b}$, Gia~Dvali$^{c,d}$, Stefan~Hofmann$^{a,e}$, Justin~Khoury$^a$, Oriol~Pujol\`as$^d$, Michele~Redi$^{d,f}$ and Andrew~J.~Tolley$^a$}
\address{$^a$Perimeter Institute for Theoretical Physics, 31 Caroline St. N., Waterloo, ON, N2L 2Y5, Canada}
\address{$^b$Dept. of Physics \& Astronomy, McMaster University, Hamilton ON, L8S 4M1,  Canada}
\address{$^c$CERN, Theory Division, CH-1211 Geneva 23, Switzerland}
\address{$^d$Center for Cosmology and Particle Physics,
New York University, New York, NY 10003 USA}
\address{$^e$NORDITA, Roslagstullsbacken 23, 106 91 Stockholm, Sweden}
\address{$^f$Institut de Th\'eorie des Ph\'enom\`enes Physiques, EPFL, CH-1015, Lausanne, Switzerland}

\begin{abstract}
We present a higher codimension generalization of the DGP scenario
which, unlike previous attempts, is free of ghost instabilities.
The 4D propagator is made regular by embedding our visible 3-brane
within a 4-brane, each with their own induced gravity terms, in a
flat 6D bulk. The model is ghost-free if the tension on the
3-brane is larger than a certain critical value, while the induced
metric remains flat. The gravitational force law ``cascades'' from
a 6D behavior at the largest distances followed by a 5D and
finally a 4D regime at the shortest scales.
\end{abstract}

\maketitle

The DGP model \cite{DGP} provides a simple mechanism to modify
gravity at large distances by adding a localized graviton kinetic
term on a codimension 1 brane in a flat 5D bulk. The natural
generalization to higher codimension,
however, is not so straightforward. On one hand these models
require some regularization due to the divergent behavior of the
Green's functions in higher codimension. More seriously, most
constructions seem to be plagued by ghost instabilities
\cite{dubov,gregshif} (see \cite{Kaloper} for related work). The
purpose of this letter is to show that both pathologies can be
resolved by embedding a succession of higher-codimension DGP
branes into each other.

\section{Scalar}
We shall focus on the codimension 2 case. As a warm-up exercise,
we consider a real scalar field with action,
\begin{equation}
S=\frac 1 2 \int \phi \left[M_6^4 \square_6 + M_5^3 \square_5
\delta(y)+ M_4^2 \square_4 \delta(y)\delta(z)\right]\phi \nonumber
\end{equation}
describing a codimension 2 kinetic term embedded into a
codimension 1 one in 6D. We will impose throughout the paper $Z_2
\times Z_2$ orbifold projection identifying $y \to -y$ and $z\to
-z$. The model possesses the two mass scales,
\begin{equation}
m_5=\frac {M_5^3} {M_4^2} \qquad {\rm and } \qquad m_6=\frac {
M_6^4} {M_5^3}~.
\end{equation}

In absence of the 4D kinetic term the propagator on the
codimension 1 brane (4-brane) is the DGP propagator \cite{DGP},
\begin{equation}
G^0(y-y')=\frac 1 {M_5^3} \int \frac {dq} {2\pi} \frac {e^{i q
(y-y')}}{p^2+q^2+2  m_6 \sqrt{p^2+q^2}}~,
\end{equation}
where $p$ is 4D momentum and $y$ the coordinate orthogonal to the
codimension 2 brane (3-brane). To find the exact 5D propagator we
can treat the 4D kinetic term (located at $y=0$) as a perturbation
and then sum the series. One finds,
\begin{align}
&G^{exact}= G_0(y-y') - M_4^2 G^0(y) p^2 G^0(-y') \cr &~~~~~~~~~+
M_4^4 G^0(y) p^4 G^0(0) G^0(-y')+\dots \cr &=G^0(y-y')- \frac
{M_4^2 p^2} {1 + M_4^2 p^2 G^0(0)} G^0(y) G^0(-y')~.
\end{align}
In particular the 4D brane-to-brane propagator is determined in
terms of the integral of the higher dimensional Green's function,
\begin{equation}
G_4^{exact}= \frac {G^0(0)} {M_4^2 G^0(0) p^2+1}~.
\end{equation}

For the case at hand,
\begin{eqnarray}
&&G^0(0)=\frac 1 {M_5^3}\int_{-\infty}^{\infty} \frac{d q}{2 \pi}
\frac{1}{p^2+q^2+2 m_6 \sqrt{p^2+q^2}}\nonumber\\
&=&  \frac{2}{\pi M_5^3}\frac{1}{\sqrt{4 m_6^2-p^2}} \tanh^{-1}
\,\left(\sqrt{\frac{2 m_6-p}{2 m_6+p}}\,\right)~.
\end{eqnarray}
For $p>2m_6$, the analytic continuation of this expression is
understood.

Remarkably, the 5D kinetic term makes the 4D propagator finite,
thereby regularizing the logarithmic divergence characteristic of
pure codimension 2 branes. In particular, when $M_5$ goes to zero
one has $G_4^{exact} \to M_6^{-4}\log(p/m_6)$, reproducing the
codimension 2 Green's function with physical cutoff given by
$m_6$.
The corresponding 4D Newtonian potential scales as $1/r^3$ at the
largest distances, showing that the theory becomes six
dimensional, and reduces to the usual $1/r$ on the shortest
scales. Its behavior at intermediate distances, however, depends
on $m_{5,6}$. If $m_5>m_6$ there is an intermediate 5D regime;
otherwise the potential directly turns 6D at a distance of order
$(m_5 m_6)^{-1/2}\log(m_5/m_6)$.\\[2mm]


\section{Gravity}
Let us now turn to gravity. In analogy with the scalar we consider
the action,
\begin{equation}
S={M_6^4\over2}\int \sqrt{-g_6} R_6+{M_5^3\over2}\int \sqrt{-g_5}
R_5 + {M_4^2\over2}\int \sqrt{-g_4} R_4\nonumber
\end{equation}
where each term represents the intrinsic curvature. This guarantees
that the model is fully 6D general covariant.

To find the propagator it is convenient to follow the same
procedure as for the scalar and sum the diagrams with insertion of
the lower dimensional kinetic term, i.e. the Einstein's tensor
${\cal E}$. For our purpose we only compute the propagator on the
3-brane. Given the higher dimensional propagator, the
brane-to-brane propagator due to the insertion of a codimension 1
term is in compact form,
\begin{eqnarray}
G_{\mu\nu\alpha\beta}^{exact}&=& G^0 \sum_{n=0}^{\infty} (M_4^2
{\cal E} G^0)^n=
G^0 [1-M_4^2 {\cal E} G^0]^{-1} \nonumber \\
&=& G^0_{\mu\nu\gamma\delta}  H^{\gamma\delta}_{\alpha\beta}~,
\label{fullgravity}
\end{eqnarray}
where $G^0_{\mu\nu\gamma\delta}$ is the 4D part of the higher
dimensional Green's function evaluated at zero. The tensor
$H^{\mu\nu}_{\alpha\beta}$ satisfies by definition,
\begin{equation}
[1-M_4^2  {\cal E} G^0]^{\mu\nu}_{\gamma\delta}
H^{\gamma\delta}_{\alpha\beta}=\frac 1 2 \left(
\delta_\mu^\alpha\delta_\nu^\beta+\delta_\mu^\beta\delta_\nu^\alpha\right).
\label{inverse}
\end{equation}
To find $H$ one can write the most general Lorentz covariant
structure compatible with the symmetries,
\begin{eqnarray}
H^{\gamma\delta}_{\alpha\beta} &=& a (\delta^\gamma_\alpha
\delta^\delta_\beta+
\delta^\delta_\alpha \delta^\gamma_\beta)+ b \eta^{\gamma\delta} \eta_{\alpha\beta} \nonumber \\
&+& c ( p^\gamma p_\alpha \delta^\delta_\beta+p^\delta p_\alpha
\delta^\gamma_\beta+ p^\gamma p_\beta
\delta^\delta_\alpha+p^\delta p_\beta \delta^\gamma_\alpha)\cr
&+&d\, p^\gamma p^\delta \eta_{\alpha\beta}+e\,
\eta^{\gamma\delta} p_\alpha p_\beta \nonumber  + f\, p^\gamma
p^\delta p_\alpha p_\beta~.
\end{eqnarray}
Requiring that this satisfies Eq. (\ref{inverse}) leads to a
system of linear equations whose solution determines the
coefficients $a, b, c, d, e, f$. Using this information one then
reconstructs the exact propagator from Eq. (\ref{fullgravity}).

It is straightforward to apply this technique to the cascading DGP.
Starting from 6D, the propagator on the 4-brane is
\cite{gregorygia},
\begin{eqnarray}
G_{MNPQ}&=&\frac 1 {M_5^3}\frac 1 {p_5^2+2 m_6 p_5}\times \nonumber \\
&\times& \left(\frac 1 2 \tilde{\eta}_{MP}\tilde{\eta}_{NQ}+\frac 1
2 \tilde{\eta}_{MQ}\tilde\eta_{NP}-\frac 1 4
\tilde{\eta}_{MN}\tilde{\eta}_{PQ}\right) \,,\nonumber \\
\tilde{\eta}_{MN}&=&\eta_{MN}+ \frac {p_M p_N} {2 m_6 p_5} ~,
\label{dgppropagator}
\end{eqnarray}
where $M, N \dots$ are 5D indices and $p_5^2=p_M p^M$.
$G_{\mu\nu\alpha\beta}^0$ is obtained by integrating the 5D
propagator with respect to the extra-momentum. To compute the
propagator on the 3-brane, then, we determine the coefficients $a,
b, c, d, e, f$ through the system of linear equations
(\ref{inverse}). One finds,
\begin{eqnarray}
a &=&- \frac 1 {2 (I_1 p^2+ 1) } \nonumber \\
b&=& \frac {I_1 p^2}{(I_1 p^2+1)(I_1 p^2-2)}\nonumber \\
c&=& -\frac {I_1}{2 (I_1 p^2+ 1)}\nonumber \\
d&=& -\frac {I_1}{(I_1 p^2+1)(I_1 p^2-2)}\nonumber \\
e &=& \frac {1}{3 (I_1 p^2+1)}- \frac {4 I_1 + 3 I_2 p^2}{3 (I_1 p^2-2)}\nonumber \\
f &=& \frac {I_2+2 I_1^2+ I_1 I_2 p^2}{(I_1 p^2+1)(I_1 p^2-2)}
\end{eqnarray}
where
\begin{eqnarray}
I_1&=&\frac 1 {m_5} \int \frac {dq} {2\pi} \frac 1 {p^2+q^2+2 m_6 \sqrt{p^2+q^2}}\nonumber \\
I_2&=&\frac 1 {2 m_6m_5} \int \frac {dq} {2\pi} \frac 1
{\sqrt{p^2+q^2}(p^2+q^2+2 m_6 \sqrt{p^2+q^2})} \nonumber
\end{eqnarray}
All these coefficients are finite showing that the regularization
is also effective for the spin 2 case.


Having determined the coefficients of the tensor $H$ the full
propagator is given by Eq. (\ref{fullgravity}). To linear order
the amplitude between two conserved sources on the brane is rather
simple,
\begin{equation}
-\frac {1}{M_4^2} \frac
{I_1}{I_1p^2+1}\left(T_{\mu\nu}T'^{\mu\nu}-\frac {I_1 p^2-1}{2 I_1
p^2-4} T T'\right)~, \label{amplitude2}
\end{equation}
and only depends on the first integral $I_1$.

The coefficient in front of the amplitude is exactly as for the
scalar however there is a non-trivial tensor structure. One
worrisome feature of this amplitude is that the relative
coefficient of $T_{\mu\nu}T'^{\mu\nu}$ and $T T'$ interpolates
between $-1/4$ in the IR and $-1/2$ in the UV. The $-1/4$ in the
IR gives the correct tensor structure of gravity in 6D and is
unavoidable because at large distances the physics is dominated by
6D Einstein term. From the 4D point of view this can be understood
as the exchange of massive gravitons and an extra-scalar. The
$-1/2$ in the UV on the other hand signals  the presence of a
ghost. This agrees with previous results \cite{gregshif,dubov}
which used a different regularization. From the 4D point of view
the theory decomposes into massive spin 2 fields and scalars.
Since the massive spin 2 gives an amplitude with relative
coefficient $-1/3$ the extra repulsion must be provided by a
scalar with wrong sign kinetic term. Separating from Eq.
(\ref{amplitude2}) the massive spin two contribution, we identify
the scalar propagator as
\begin{equation}
G_{ghost}= \frac{1}{6M_4^2} \frac {I_1} {I_1 p^2-2 }~.
\end{equation}
This propagator has a pole with negative residue therefore it
contains a localized (tachyonic) ghost mode in addition to a
continuum of healthy modes.\\

\section{Ghost free theory}
To clarify the origin of the ghost it is illuminating to consider
the decoupling limit studied in \cite{lpr}. This will allow us to
show how a healthy theory can be obtained by simply introducing
tension on the lower dimensional brane while retaining the
intrinsic geometry flat.

In the 6D case the decoupling limit \cite{lpr} corresponds to
taking $M_5,M_6 \to \infty$ with $\Lambda_s\equiv (m_6^2
M_5^{3/2})^{2/7}=(M_6^{16}/M_5^9 )^{1/7}$ finite. In this limit,
the physics on the 4-brane admits a local 5D description, where
only the non-linearities in the helicity 0 part of the metric are
kept, and are suppressed by the scale $\Lambda_s$. The effective
5D lagrangian is given by
\begin{eqnarray}\label{5Daction}
L_5 &=& {M_5^3\over4} \, h^{MN} (\EE h)_{MN} - 3 M_5^3 (\partial
\pi)^2 \lp( 1 + {9\over 32m_6^2} \,\Box_5\pi  \rp) \cr %
&+&\delta(z)\lp( {M_4^2\over4} \, \WT h^{\mu\nu} (\EE \WT
h)_{\mu\nu} + \WT h^{\mu\nu} T_{\mu\nu} \rp) ~,
\end{eqnarray}
where $(\EE h)_{MN}=\Box h_{MN}+\dots$ is the linearized Einstein
tensor, $M, N \dots$ are 5D indices and $\mu,\nu\dots$ are four
dimensional. We have rescaled $\pi$ and $h_{\mu\nu}$ so that they
are dimensionless, and the physical 5D metric is
\begin{equation}\label{physical}
  \widetilde h_{MN}=h_{MN}+\pi\;
  \eta_{MN}~.
\end{equation}
The first line of \eqref{5Daction} is the 5D version of the `$\pi$
Lagrangian' introduced in \cite{lpr} for the DGP model. In
addition to this, we have the localized curvature term on the
3-brane, which depends on 4D physical metric $\WT h_{\mu\nu}$.
This introduces a kinetic mixing between $\pi$ and the 5D metric.

We now take a further step and compute the boundary effective
action valid on the 3-brane. At the quadratic order by integrating
out the fifth dimension the 5D kinetic term of $\pi$ produces a 4D
``mass term'' $\sim M_5^3 \sqrt{-\Box_4}$ while the Einstein
tensor gives rise to Pauli-Fierz (PF) structure for $h_{\mu\nu}$
on the boundary\footnote{The massless 5D graviton decomposes in a
continuum of massive 4D gravitons. Therefore in the unitary gauge
the boundary effective action will have PF structure.},
\begin{eqnarray}\label{boundaryEffAct}
  L_4&=& - {M_5^3\over2}\; h^{\mu\nu}\sqrt{-\Box_4}\lp( h_{\mu\nu} - h \eta_{\mu\nu} \rp)
-  6 M_5^3\, \pi \sqrt{-\Box_4} \,\pi \cr %
&+&   {M_4^2 \over4}\; \WT h^{\mu\nu}(\EE \widetilde h)_{\mu\nu} %
  + \WT h^{\mu\nu} \, T_{\mu\nu}   ~,
\end{eqnarray}
where $h_{\mu\nu}$ and $\pi$ now denote the 5D fields evaluated at the
3-brane location.

In terms of the physical metric,  \eqref{boundaryEffAct} takes the
form,
\begin{eqnarray}\label{boundaryEffAct2}
  L_4&=  &  - {M_5^3\over2}\; \widetilde  h^{\mu\nu}\sqrt{-\Box_4}\lp(  \widetilde h_{\mu\nu} -
     \widetilde h \eta_{\mu\nu} \rp) %
     - 3 M_5^3\, \pi \sqrt{-\Box_4} \, \widetilde  h \cr
  &  +&{M_4^2\over4} \; \widetilde h^{\mu\nu}(\EE \widetilde h)_{\mu\nu}
 + \widetilde h^{\mu\nu} \, T_{\mu\nu} ~.
\end{eqnarray}
Note that the kinetic term for $\pi$ is completely absorbed by that of $\WT h_{\mu\nu}$ and
only a cross term between $\pi$ and $\WT h_\mu^\mu$ remains.

From this it is straightforward to show the presence of a ghost.
The scalar longitudinal component of $h_{\mu\nu}$ acquires a
positive kinetic term by mixing with the graviton \cite{nima,lpr}.
By taking
\begin{equation}
\widetilde{h}_{\mu\nu}=\widehat{h}_{\mu\nu}+\phi \, \eta_{\mu\nu}+
\frac {\partial_{\mu}\partial_{\nu} } {m_5 \sqrt{-\Box_4}}\,\phi~,
\end{equation}
one finds that there are in fact two 4D scalar modes whose kinetic
matrix in the UV is
\begin{equation}
{3\over2} M_4^2 \; \Box_4 \, %
\left(\begin{array}{cc} %
1 & 1 \\
1 & 0
\end{array} \right)
\label{matrix}
\end{equation}
which has obviously a negative eigenvalue corresponding to a ghost.


Having understood the origin of the ghost we are now ready to show
how to cure it. To achieve this we clearly need to introduce a
positive localized kinetic term for $\pi$. This can arise from extrinsic
curvature contributions. The simplest and most natural choice is to
put a tension $\Lambda$ on the 3-brane. This produces extrinsic
curvature while leaving the metric on the brane flat since the
tension only creates a deficit angle.

The solution to the 5D equations following
from \eqref{5Daction} for a 3-brane with tension $\Lambda$ is
\cite{dws}
\begin{equation}\label{background}
\pi^{(0)} = {\Lambda\over6M_5^3} \, |z|\;,\quad\quad
h^{(0)}_{\mu\nu}= -{\Lambda\over6M_5^3}\, |z| \, \eta_{\mu\nu}~.
\end{equation}
This is an exact solution including the non-linear
terms for $\pi$ -- they vanish identically for this profile. The
background corresponds to a locally flat 6D bulk with deficit angle
$\Lambda/M_6^4$ and with flat 4D sections.

The crucial point is that on this background the $\pi$ lagrangian
acquires contributions from the non-linear terms. These can be
found considering the perturbations,
\begin{align}\label{fluct}
  \pi&=\pi^{(0)}(z)+\delta\pi(z,x^\mu)\,, \cr
h_{\mu\nu}&=h^{(0)}_{\mu\nu}(z)+ \delta h_{\mu\nu}(z,x^\mu)\,, \cr
T_{\mu\nu}&=-\Lambda \eta_{\mu\nu} + \delta T_{\mu\nu}~.
\end{align}
Plugging \eqref{fluct} in \eqref{5Daction} and dropping $\delta$,
one obtains at quadratic order (up to a total derivative),
\begin{eqnarray}
\delta L_5 &=& {27\over 4}{M_5^3\over
m_6^2}\lp(\partial_M\pi\partial_N\pi\;\partial^M\partial^N\pi^{(0)}-
(\partial_M\pi)^2\,\Box_5\pi^{(0)}\rp) \cr %
&=& -{9\over4}{\Lambda\over m_6^2}\;\delta(z) \;(\partial_\mu
\pi)^2~. \label{addition}
\end{eqnarray}
This is a localized kinetic term for $\pi$ that contributes to the
4D effective action with a healthy sign when $\Lambda>0$. Therefore
for large enough $\Lambda$ the kinetic matrix for the 2 scalars (\ref{matrix})
becomes positive and the ghost is absent.

This can also be seen  by computing the one particle exchange
amplitude. With the addition of (\ref{addition}), the effective
4D equations are
\begin{align}
M_4^2\,({\EE \WT h})_{\mu\nu} &- 2 M_5^3 \sqrt{-\Box_4} \lp( \WT
h_{\mu\nu} - \WT h \, \eta_{\mu\nu}
 \rp)\nonumber \\
& = - 2 \, T_{\mu\nu}  + 6 M_5^3 \sqrt{-\Box_4} \,\pi
\,\eta_{\mu\nu} ~, \label{einst5} \\
{3 \Lambda \over 2 m_6^2 }  \; \Box_4 \pi&= M_5^3\, \sqrt{-\Box_4}
\, \widetilde  h ~. \label{pi}
\end{align}
Using the Bianchi identities and the conservation of $T_{\mu\nu}$,
the double divergence of \eqref{einst5} leads to
\begin{equation}\label{dd}
M_5^3 \sqrt{-\Box_4} \lp( (\EE \WT h)_\mu^\mu + 6 \Box_4 \pi
\rp)=0~,
\end{equation}
where we have used that $(\EE \WT
h)_\mu^\mu=2(\partial^\mu\partial^\nu \WT h_{\mu\nu}-\Box_4 \WT
h)$.
On the other hand, the trace of \eqref{einst5} in conjunction with
\eqref{pi} and \eqref{dd}, leads to
\begin{equation}\label{pi4}
 M_4^2 \, \OO_{(\pi)} \; \pi=  - 2 T_\mu^\mu ~, 
\end{equation}
where $\OO_{(\pi)}\equiv \lp[9\lp(\Lambda/ m_6^2 M_4^2\rp)-6\rp]
\Box_4 - 24 \, m_5 \sqrt{-\Box_4} $.\\[-2mm]

Combining \eqref{einst5}, \eqref{dd} and \eqref{pi4}, one derives
that the physical metric is, up to pure gauge terms,
\begin{equation}\label{resultTension}
\widetilde h_{\mu\nu}= {-2\over M_4^2} \Biggr\{ %
{1\over\OO} \lp(T_{\mu\nu}-{T\over3} \eta_{\mu\nu}\rp) +%
{1\over \OO_{(\pi)}} \, T \;\eta_{\mu\nu} \Biggr\} ~, %
\end{equation}
where $\OO=\Box_4-2\, m_5 \sqrt{-\Box_4}$. The tensor structure of
the amplitude interpolates between $-1/4$ in the IR and
$$
-{1\over3}+ {1\over6}\lp({3\over 2}{\Lambda\over M_4^2 m_6^2} -
1\rp)^{-1}
$$
in the UV.
The amplitude above corresponds to the exchange of massive spin 2
fields and a scalar obeying Eq. (\ref{pi4}). The (DGP-like)
kinetic term for the scalar is positive as long as,
\begin{equation}\label{lambdaMin}
\Lambda > {2\over 3} M_4^2 m_6^2 ~.%
\end{equation}
In this regime we see that the localized ghost disappears and the
scalar sector is composed of a healthy resonance.

In the limit we are considering the tension required is consistent
with having six non compact dimensions. Indeed, requiring that the
deficit angle in the bulk is less than $2\pi$ leads to $\Lambda<
2\pi M_6^4$. It follows that,
\begin{equation}\label{window}
3\pi M_5^6  > {M_4^2 \,M_6^4 }~,
\end{equation}
which is always satisfied in the $5D$ decoupling limit and
displays the necessity of the induced term on the codimension 1
brane. Moreover the condition above is equivalent to having $m_6 <
m_5$, which suggests that in order to avoid the ghost one should
cascade from the highest dimension down to 4D 'step by step'.\\[-3mm]

From a phenomenological point of view, observations require that
$m_5\lesssim H_0$, the present Hubble scale. The most interesting
possibility is when the $6D$ crossover scale is larger but of
similar order. Assuming that the formulas above can be
extrapolated in this regime for a Planckian $M_4$ this implies
that $M_5$ is of order $10 MeV$ and $M_6 \sim m eV$. The latter
also sets the scale of $\Lambda$.\\[-1mm]

\section{Discussion}\vspace{-4mm}
In this note we have presented a six dimensional DGP model with
cascading localized kinetic terms. The model interpolates between
a 6D behavior at large distances and a 4D one at short distances
with an intermediate 5D regime. The kinetic terms regularize the
divergent codimension 2 behavior. The model is ghost-free at least
for a certain range of parameters if on the codimension 2 brane
there is a large enough tension.\\[-5mm]

We have left several questions for future study. At linear-level
the tensor structure of the graviton propagator is inconsistent
with observations. In the context of DGP this was shown not to be
a problem because the non-linearities restore the correct tensor
structure \cite{vainshtein}.
A hint of a similar phenomenon in the present model is given by
the longitudinal terms of the graviton propagator
\eqref{fullgravity}. These are singular when the mass parameters
$m_{5,6}$ vanish and give large contributions to nonlinear
diagrams. In fact, we expect a `double' Vainshtein effect. For
dense enough sources, the non-linearities should first decouple
the extra 5D scalar mode restoring 5D behavior followed by another
step to 4D.
Another important direction to study is cosmology. The model has
the intriguing codimension 2 feature that tension does not curve
the space. This obviously of interest for the cosmological
constant problem.\\[-4mm]

\emph{Acknowledgements} We thank Gregory Gabadadze for useful
discussions. This work is supported in part by NSF grant
PHY-0245068 and by the David and Lucile Packard Foundation
Fellowship for Science and Engineering (GD and MR), by DURSI under
grant 2005 BP-A 10131 (OP), and by NSERC and MRI (JK, AJT, CdR and
SH).

\end{document}